\documentclass[manuscript]{aastex}

\usepackage{color}
\usepackage{ulem}

\usepackage{lineno}
\usepackage{multirow}
\linenumbers
\slugcomment{To be submitted to ApJ}
\shorttitle{An Anisotropy study of UHECRs}
\shortauthors{The Telescope Array Collaboration}

\begin{document}
\title{SEARCH FOR CORRELATIONS OF THE ARRIVAL DIRECTIONS OF ULTRA-HIGH 
ENERGY COSMIC RAY WITH EXTRAGALACTIC OBJECTS AS
OBSERVED BY THE TELESCOPE ARRAY EXPERIMENT}

 \author{
T.~ABU-ZAYYAD$^{1}$, 
R.~AIDA$^{2}$, 
M.~ALLEN$^{1}$, 
R.~ANDERSON$^{1}$, 
R.~AZUMA$^{3}$, 
E.~BARCIKOWSKI$^{1}$, 
J.W.~BELZ$^{1}$,
D.R.~BERGMAN$^{1}$, 
S.A.~BLAKE$^{1}$, 
R.~CADY$^{1}$
% ...
\\~
}
\affil{
$^1$High Energy Astrophysics Institute and Department of Physics and Astronomy, University of Utah, Salt Lake City, Utah, USA\\
$^2$University of Yamanashi, Interdisciplinary Graduate School of Medicine and Engineering, Kofu, Yamanashi, Japan\\
$^3$Graduate School of Science and Engineering, Tokyo Institute of Technology, Meguro, Tokyo, Japan\\
% ...
}

\begin{abstract}
\linenumbers

We search for correlations between 
positions of extragalactic objects and arrival directions of
Ultra-High Energy Cosmic Rays (UHECRs)
with primary energy $E \ge 40$~EeV as observed by the surface detector 
array of the Telescope Array (TA) experiment during the first 
40 months of operation.
We examined several public astronomical object catalogs, including 
the Veron-Cetty and Veron catalog of active galactic nuclei.
We counted the number of TA events correlated with objects in each
catalog as a function of three parameters: the maximum angular separation between
a TA event and an object, the minimum energy of the events, 
and the maximum redshift of the objects.
We determine combinations of these  parameters which maximize 
the correlations, and calculate the chance probabilities
of having the same levels of correlations from an isotropic 
distribution of UHECR arrival directions.
No statistically significant correlations
are found when penalties for scanning over the above 
parameters and for searching in several catalogs are taken into account.
%
\end{abstract}


\section{INTRODUCTION}


Clarifying the origin of Ultra-High Energy Cosmic Rays
(UHECRs) is one of the most important unsolved
problems in modern astrophysics~\citep[e.g.][]{kotera}.
It is generally thought that cosmic rays with energies greater than 
10$^{18}$~eV (1~EeV) are of 
extragalactic origin because the Galactic magnetic fields are not
strong enough to confine them. 
Indeed, no apparent anisotropy in arrival directions 
of UHECRs along the Galactic plane has been found.
On the other hand, a steepening in the energy spectrum
of UHECRs at around 50~EeV is observed by the High Resolution Fly's Eye (HiRes) 
experiment and the Telescope Array (TA) experiment~\citep{HiRes_ensp,TASD_ensp},
and also by the Pierre Auger Observatory 
in a similar energy region~\citep{Auger_ensp,Auger_ensp2}.
This can be explained as a consequence of the cosmic ray energy 
losses due to interactions with the Cosmic Microwave Background 
(CMB), as predicted by ~\citep{GZK1}, and ~\citep{GZK2}.

In this case, we expect that most of the observed cosmic rays of the 
highest energies originate from sources within the GZK horizon 
($\sim$100~Mpc), and a correlation
between nearby objects and arrival directions of cosmic 
rays is expected.
The UHECRs are deflected 
by the Galactic and extragalactic magnetic fields on 
their way to Earth.
The deflection angles are determined by the
particle charges, source distances, and strength of the 
magnetic fields.
For example, in case of a proton arriving from a 100~Mpc distance
through a random extragalactic magnetic field 1~nG 
and correlation length of $\sim$1~Mpc, 
the expected deflection angle is $3-5^{\circ}$ for 100~EeV 
(and less than $15^{\circ}$ for 40~EeV)
 using the existing magnetic field estimates
\citep{Han,Sun,Psirkov,Kronberg}.

The TA experiment observes UHECRs in the northern hemisphere 
using a Surface Detector (SD) array~\citep{SD} of $\sim$ 700~km$^2$ area
located in Millard County, Utah, USA (39.3$^{\circ}$~N, 112.9$^{\circ}$~W).
Three Fluorescence Detector (FD) stations~\citep{ta_fd,jnm} 
surround the SD array~\citep{sogio} and view the atmosphere above it.
The SD array consists of 507 SDs installed on a square grid with 1.2~km spacing, 
and measures particles from Extensive Air Showers (EASs) at ground level.
The energy and the arrival direction of a 
primary particle
are determined from observed energy deposits as a function of distance from the shower core in the SDs 
and the arrival time distribution of the EAS particles.
The test operation of the SD array began in March 2008, and
the full SD array has been operational with a uniform trigger criteria 
from May 11, 2008.
%
%
The present analysis uses only the events detected by the SD array 
because this data set has the greatest statistics than that by the FDs.



Assuming the sources have the same intrinsic UHECR luminosities, the
arrival directions of higher energy cosmic rays 
from nearby sources are expected to correlate better with the source positions. 
We search for the correlations between the TA events and
objects in catalogs by changing three parameters:
the minimum energy of the cosmic ray events, $E_{\rm min}$, 
the separation angle, $\psi$, between the cosmic ray arrival direction and 
the object, and the maximum redshift, $z_{\rm max}$, of the objects.
A similar approach has been taken in the analyses by the 
Pierre Auger Observatory~\citep{AugerScience,AugerAstropartphys,AugerUpdate}
and by the HiRes experiment~\citep{HiRes_VCV}
using the VCV catalog of 12th edition~\citep{VCV2006}.

%
%
As putative sources of UHECR, we 
examine the objects in the 13th edition of the VCV catalog~\citep{VCV}.
This catalog is a compilation of several surveys made under
different conditions such as Field Of Views (FOVs), observation periods, etc.
It does not represent a
homogeneous sample of Active Galactic Nuclei (AGNs), 
and its degree of completeness is unknown~\citep{VCV}.
In addition, we have investigated unbiased data sets from different 
measurements, namely, radio: the third Cambridge catalog of radio sources catalog (3CRR)~\citep{3CRR},
infrared: the 2MASS (the Two Micron All-Sky Survey) redshift survey catalog (2MRS)~\citep{2MRS}, 
X-Ray: Swift BAT (Burst Alert Telescope) 58-Month hard X-ray survey catalog (SB-58M)~\citep{SwiftBAT} 
and 60-Month AGN survey catalog (SB-AGN)~\citep{SwiftBATAGN}, 
and Gamma-ray: 2nd Fermi large area telescope AGN catalog (2LAC)~\citep{2FGL}.
In each catalog, we select only those objects that have redshift information.
In the case of the 2LAC catalog, this criterion reduces the number of  
objects by $\sim$50\%. 
%


The paper is organized as follows.
The observation status of the SD array and qualities of 
reconstructed events are briefly described in Section~\ref{SD_data}. 
The details of the parameter scanning in the correlation searches
using the object catalogs are given in Section~\ref{correlation}, 
and the results are described in Section~\ref{results}.
We also investigated penalties for the multi-catalog scanning 
in Section~\ref{discussion}.
The conclusions from this analysis are in Section~\ref{Conclusion}.

\section{SD DATA}\label{SD_data}

In this work we use the SD air shower events observed in the first
40-month run of TA from May 2008 through September 2011.
These events are triggered by a three-fold coincidence of adjacent SDs
within $8\mu$s \citep{SD}.  
%
%



The details of SD event reconstruction are described 
elsewhere~\citep{icrc_tasp,TASD_ensp}.
First, the shower geometry including the arrival direction is 
obtained using the time differences between the observed signals at each SD.
Next, the precise shower geometry and the lateral distribution of 
shower particles 
are determined using the observed energy deposit in each SD.
Finally, the primary energy is determined from 
the lateral distribution. 
The overall energy scale of the SD events is fixed by calibration with
the FD energy scale using a hybrid event set as described in the 
reference~\citep{tsunesada}.
The systematic uncertainty in energy determination is $22\%$.

The data quality cuts for the reconstructed events
are the same as in the previous TA analysis papers~\citep{TAanisoApJ,TASD_ensp}.
The events are cut if the zenith angle is greater than 45$^{\circ}$
and/or the core position is within 1200~m of the SD array boundary.
The EAS reconstruction efficiency under these criteria is greater than 98\% 
including the duty cycle of the SD array for $E>10$~EeV~\citep{TASD_ensp,SD}.
The accuracy in arrival direction determination is $1.5^{\circ}$ and 
the energy resolution is better than 20\% in this energy range.

\begin{figure}
\begin{center}
\includegraphics[angle=0,keepaspectratio,scale=0.4]{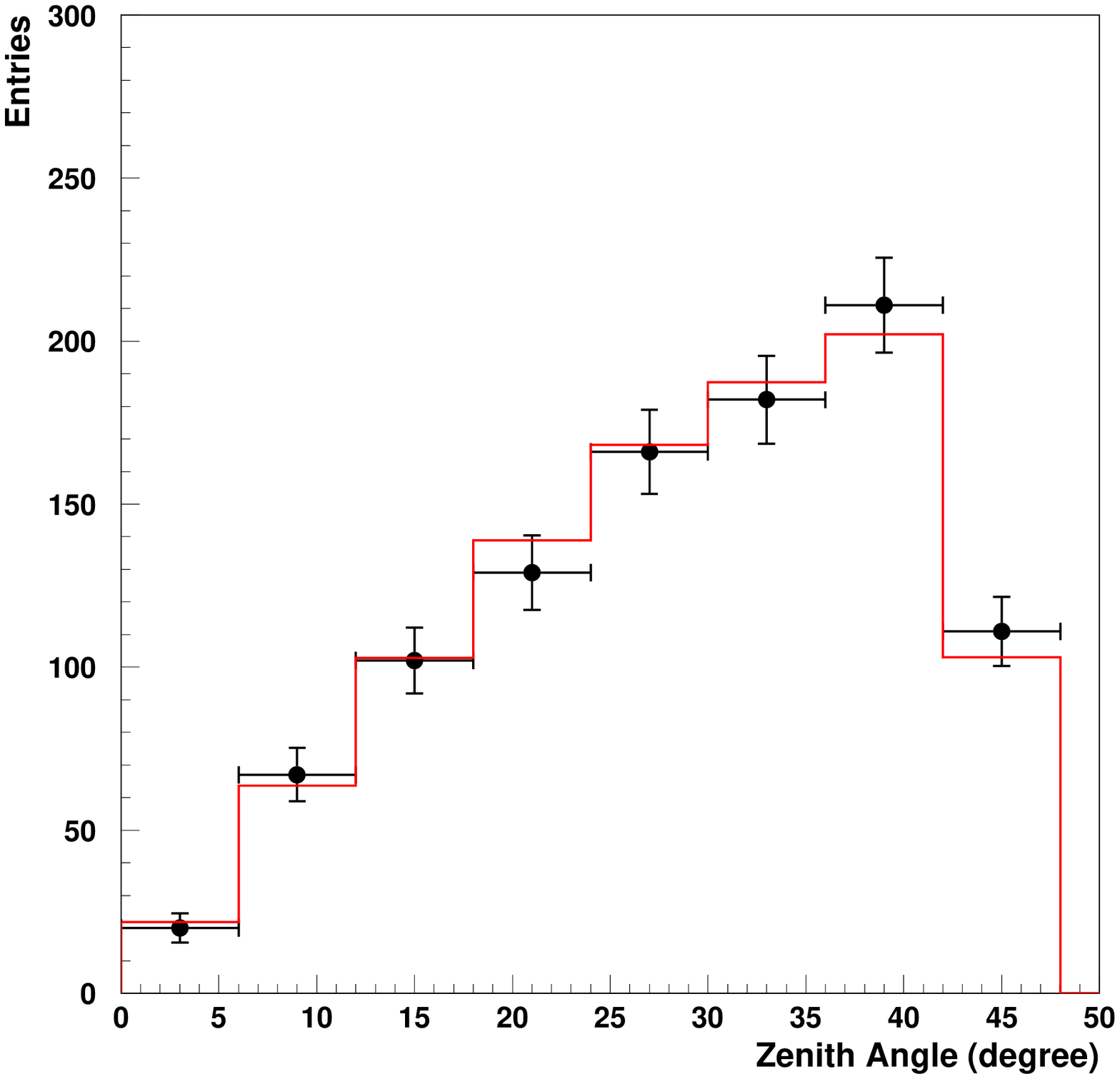}
\includegraphics[angle=0,keepaspectratio,scale=0.4]{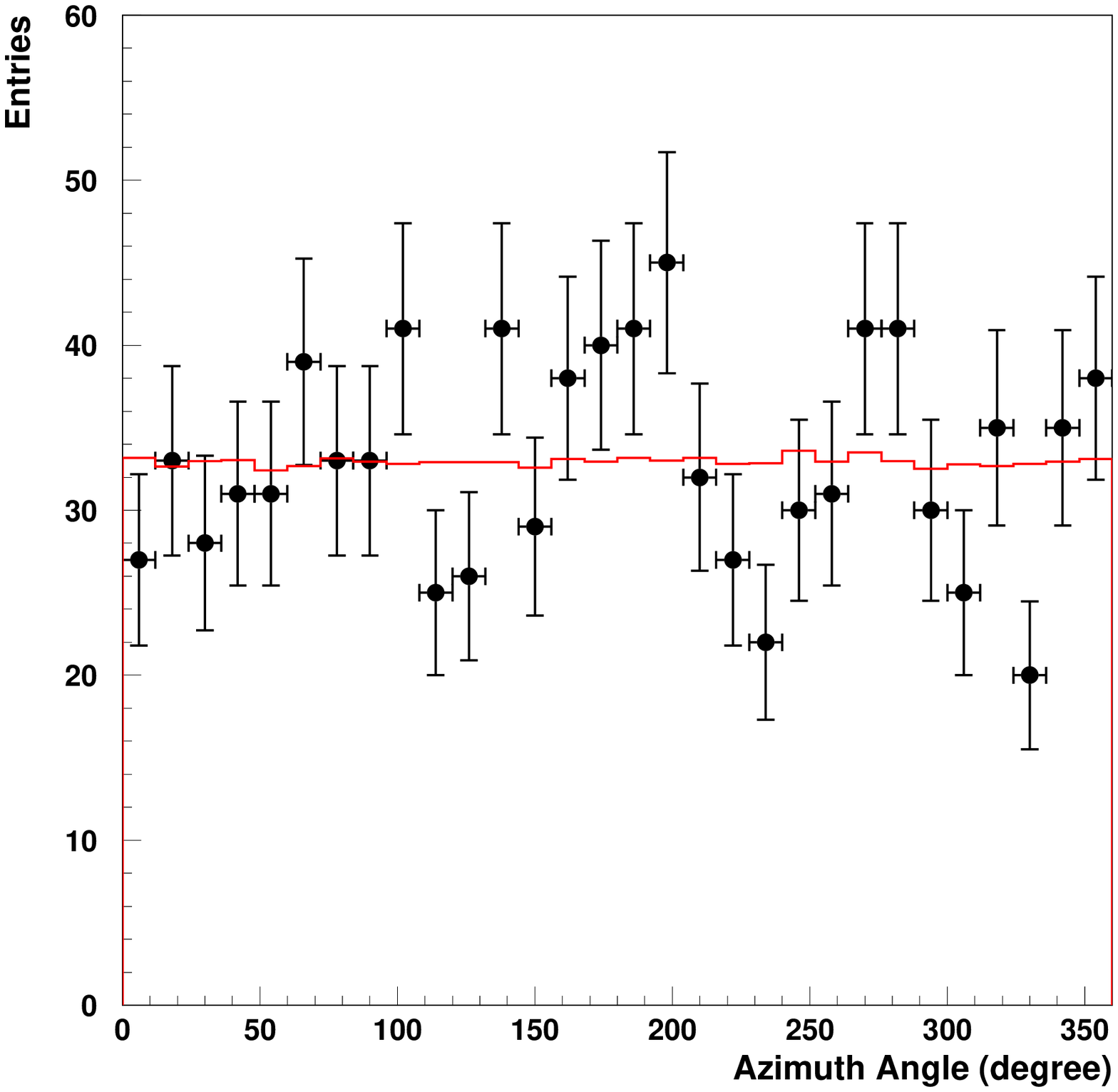}
\includegraphics[angle=0,keepaspectratio,scale=0.4]{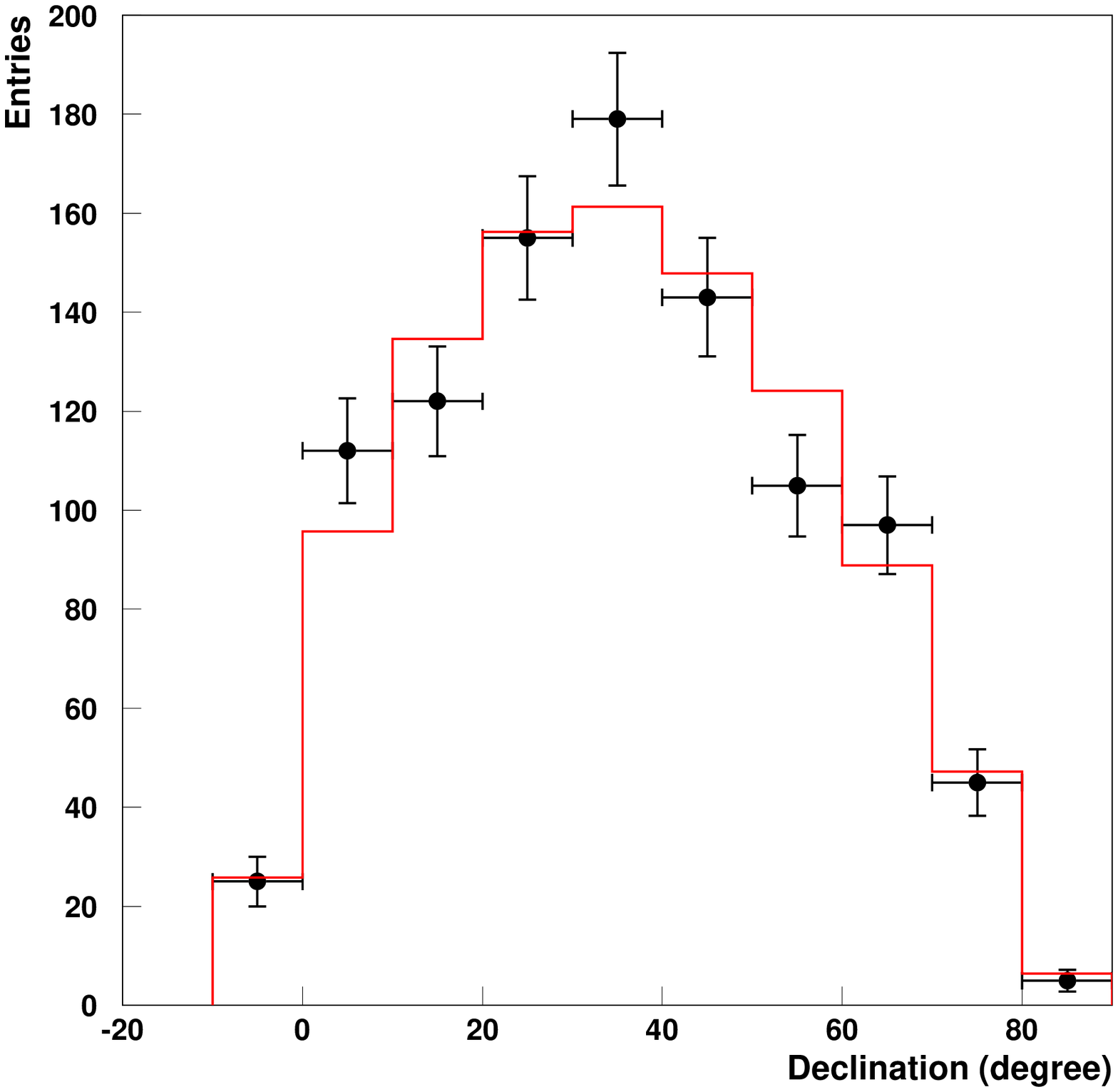}
\includegraphics[angle=0,keepaspectratio,scale=0.4]{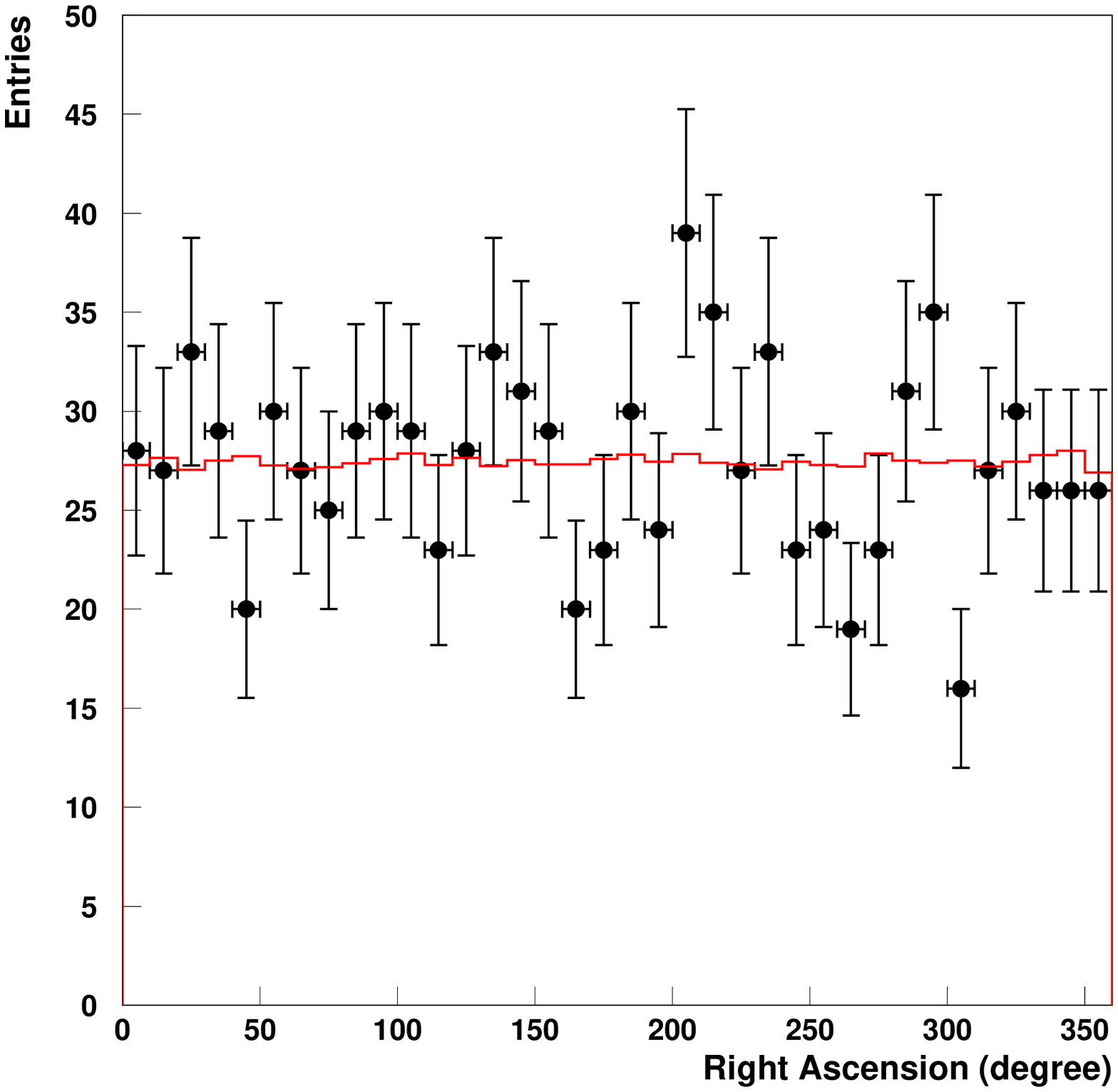}
\end{center}
\caption{The distribution of observed data (plot) and the simulated data with the geometrical acceptance (histgram) with
the energy $>$ 10~EeV.
The top left: zenith angle, the top right: azimuth angle, the bottom left: declination, and the bottom right: right accension.}\label{zenazi}
\end{figure}

The number of events remaining after reconstruction and quality cuts is
988 for $E\ge$10~EeV, 57 for $E\ge$40~EeV, and 3 
for $E\ge100$~EeV.
From our Monte-Carlo studies including the full detector response
simulations, we confirmed that the acceptance of the SD array
is fully geometrical, i.e., independent of the arrival direction up to 
$\theta = 45^{\circ}$ for showers with energies greater than 10~EeV~\citep{TAanisoApJ,TASD_ensp,SD}.
We also confirmed that the arrival direction distribution of the 
observed events in the horizontal coordinates and the equatorial
coordinates are consistent with 
large scale isotropy shown in Figure~\ref{zenazi}.
In this analysis, we use the geometrical acceptance to generate random events 
for reasons of computational efficiency.
The total exposure of the SD array in the first 40 months of operation
is $3.1\times10^3~{\rm km^2}~{\rm sr}~{\rm yr}$ including the quality cuts.

\section{CORRELATION ANALYSIS}\label{correlation}

\subsection{OBJECT CATALOGS}
We use the catalogs of extragalactic objects resulting from measurements
 as listed in Table~\ref{catalog_table}. 
In several catalogs, the objects near the Galactic plane are excluded 
to avoid incompleteness from the experimental limitation by the authors of each catalog.
We also exclude the observed SD events in the corresponding regions.

The target objects and the cut criteria in the each catalog are summarized below.
These criteria (e.g. significance level) were chosen by the authors of the each catalog.
The 3CRR catalog contains radio galaxies detected at 178~MHz 
with fluxes greater than 10~Jy~\citep{3CRR}. 
Objects in the direction of the Galactic disk 
($|b|<10^{\circ}$)
were not included. 
%
The 2MRS~\citep{2MRS} catalog is derived from the 
2MASS observation with detection range between $1-2~\mu$m and K$_{\rm s}\le 11.75$ magnitude.
This catalog also loses completeness near the Galactic plane, 
so the authors of the catalog excluded regions with 
$|b|<5^{\circ}$ for $30^{\circ} \le l \le 330^{\circ}$ and  $|b|<8^{\circ}$ otherwise.
The SB-58M catalog consists of objects 
which were detected with a significance greater than $4.8\sigma$ in
the energy range of $14-195$~keV in the first 
58 months of observation by Swift BAT. 
We select the extragalactic objects in this catalog for this work.
The catalog of SB-AGN contains AGNs with at least 
$5\sigma$ significance in the energy range of
$15-55$~keV in the first 60 months of observation by Swift BAT.
The 2LAC~\citep{2FGL} data set consists of AGNs detected
with at least $4\sigma$ significance in the 
energy range of 100~MeV$-$100~GeV in the first 24 months of observation by Fermi-LAT.
The region of the Galactic disk $|b|<$10$^{\circ}$ is cut away. 
We also examine the VCV catalog which is a compilation of 
several AGN surveys. 
The number of objects and the SD events after the cuts applied in each case
 are given in Table~\ref{arrivaldirection_cut}.

\subsection{METHODS}
For a given set of parameters $(E_{\rm min}, \psi, z_{\rm max})$, there are
$N$ events with energies $E \ge E_{\rm min}$.
We can count the number of events, $k$, out of $N$ 
which are correlated with objects in a catalog with redshifts 
$z \le z_{\rm max}$ and within the angular separation $\psi$.
We can calculate the chance probability, $P$, 
that $k$ or more correlated events are found from an 
isotropic UHECR flux under the same conditions.
We carry out a parameter scan in 
$(E_{\rm min}, \psi, z_{\rm max})$ space to find the set of parameters
which maximizes the correlation between the TA events and the 
catalog objects, i.e., minimizes $P$. 
To determine the probability, $P$, we
first obtain the probability, $p$, that a random event is correlated
with at least one object by chance for a given $(\psi, z_{\rm max})$.
We generate 10$^4$ random events to obtain the probability, $p$,
in the same experimental region of the each catalog.

Then $P$ can be obtained as a cumulative binomial probability:
\begin{eqnarray}
P= \sum^{N}_{j=k}C^{N}_{j}p^{j}(1-p)^{N-j}.
\label{cumulative_binomial_probability}
\end{eqnarray}

The scan over parameters was performed as follows. 
The value of $E_{\rm min}$ is set by 
the energy of the $N$-th highest energy event.
We scan over all values of $N$ such that $E_{\rm min}$
is greater than 40~EeV.
Note that this energy is less than the energy (50~EeV) at which
the TA energy spectrum begins to fall off steeply~\citep{TASD_ensp}.
We set the upper boundary of the parameter $z_{\rm max}$ as
0.03, which corresponds to the distances smaller than 120~Mpc.
This is comparable to the GZK horizon.
The selected step size of $z_{\rm max}$ is
$0.001$, which is
the typical accuracy in the redshift measurements. 
The separation angle, 
$\psi$, is varied from $1^{\circ}$ to $15^{\circ}$.
The maximum search window of $\psi=15^{\circ}$ is selected as 
appropriate for lower energy events ($\sim$40~EeV) arriving from the distance of 100~Mpc.
The selected step size in $\psi$ is chosen as $0.1^{\circ}$ for $\psi < 8^{\circ}$ 
and $1^{\circ}$ for $8^{\circ} \le \psi \le 15^{\circ}$ .
The parameter ranges and step sizes are summarized in Table~\ref{parameter_table}.

The minimum $P$ obtained from this procedure does not represent the 
correlation probability directly, 
because the parameter scanning enhances the correlation probability 
artificially~\citep{penalties}.
Therefore, a penalty for parameter scanning should be evaluated 
and the true probability of correlation must include this penalty. 
This will be described in Section~\ref{results}.

\section{RESULTS}\label{results}

\subsection{RESULTS OF THE PARAMETER SCAN ANALYSIS}

\begin{figure}
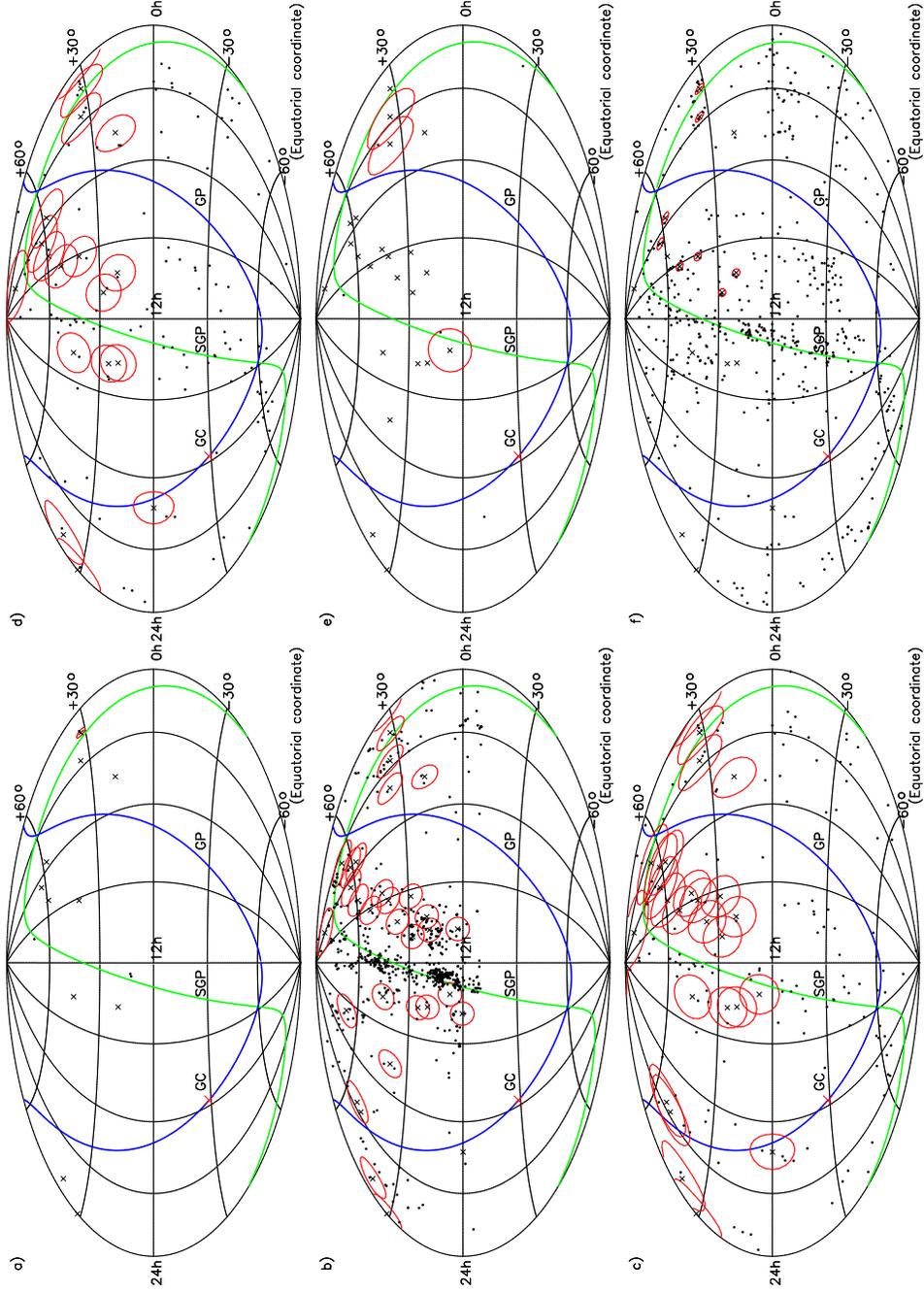

\begin{center}
\includegraphics[angle=0,keepaspectratio,scale=0.39,angle=0]{swiftbatagn.eps}
\includegraphics[angle=0,keepaspectratio,scale=0.39,angle=0]{2fgl.eps}
\includegraphics[angle=0,keepaspectratio,scale=0.39,angle=0]{vcv.eps}
\includegraphics[angle=0,keepaspectratio,scale=0.39,angle=0]{3crr.eps}
\includegraphics[angle=0,keepaspectratio,scale=0.39,angle=0]{2mrs.eps}
\includegraphics[angle=0,keepaspectratio,scale=0.39,angle=0]{swiftbat.eps}
\end{center}
\caption{Arrival directions of observed UHECR with the objects of the each catalog
(a: 3CRR, b: 2MRS, c: SB-58M, d: SB-AGN, e: 2LAC, and f: VCV).
Dots: catalog objects,
x: arrival direction of observed cosmic rays,
Circle: window around cosmic ray events, 
GC: Galactic Center, 
GP: Galactic Plane, 
SGP: Super Galactic Plane.
}\label{swiftbatagn}
\end{figure}
\begin{figure}
\begin{center}
\includegraphics[angle=0,keepaspectratio,scale=0.4]{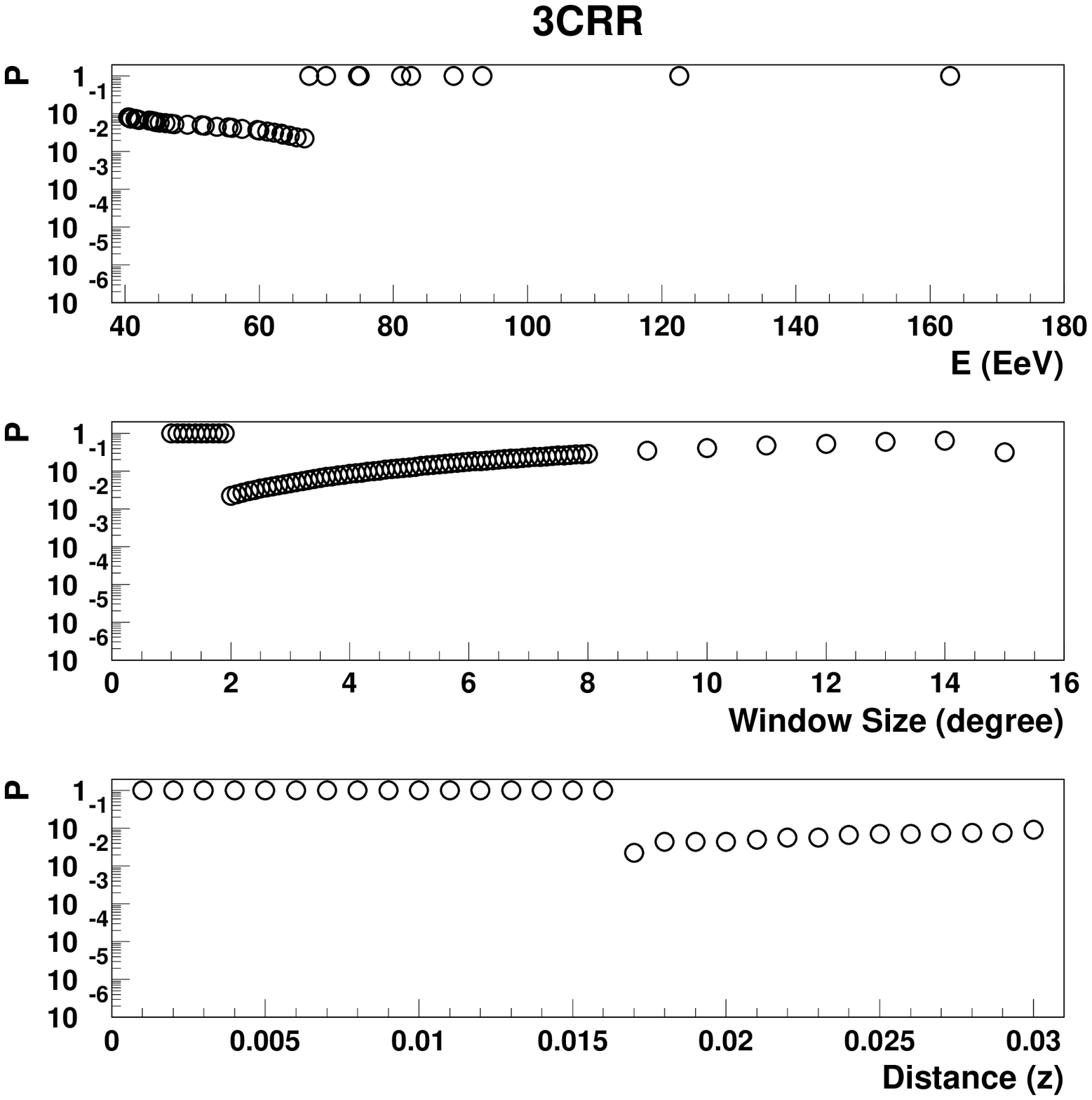}
\includegraphics[angle=0,keepaspectratio,scale=0.4]{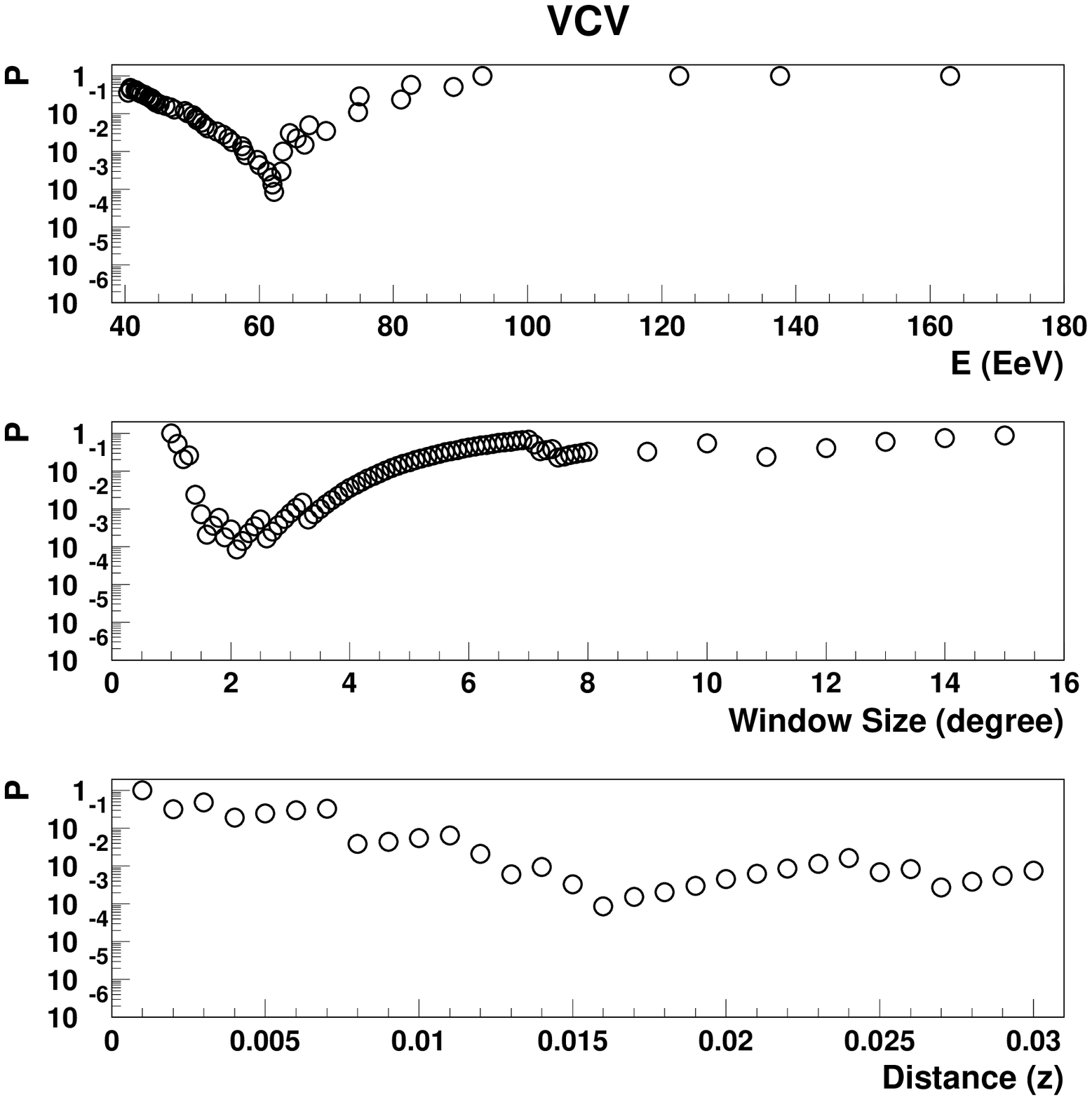}
\includegraphics[angle=0,keepaspectratio,scale=0.4]{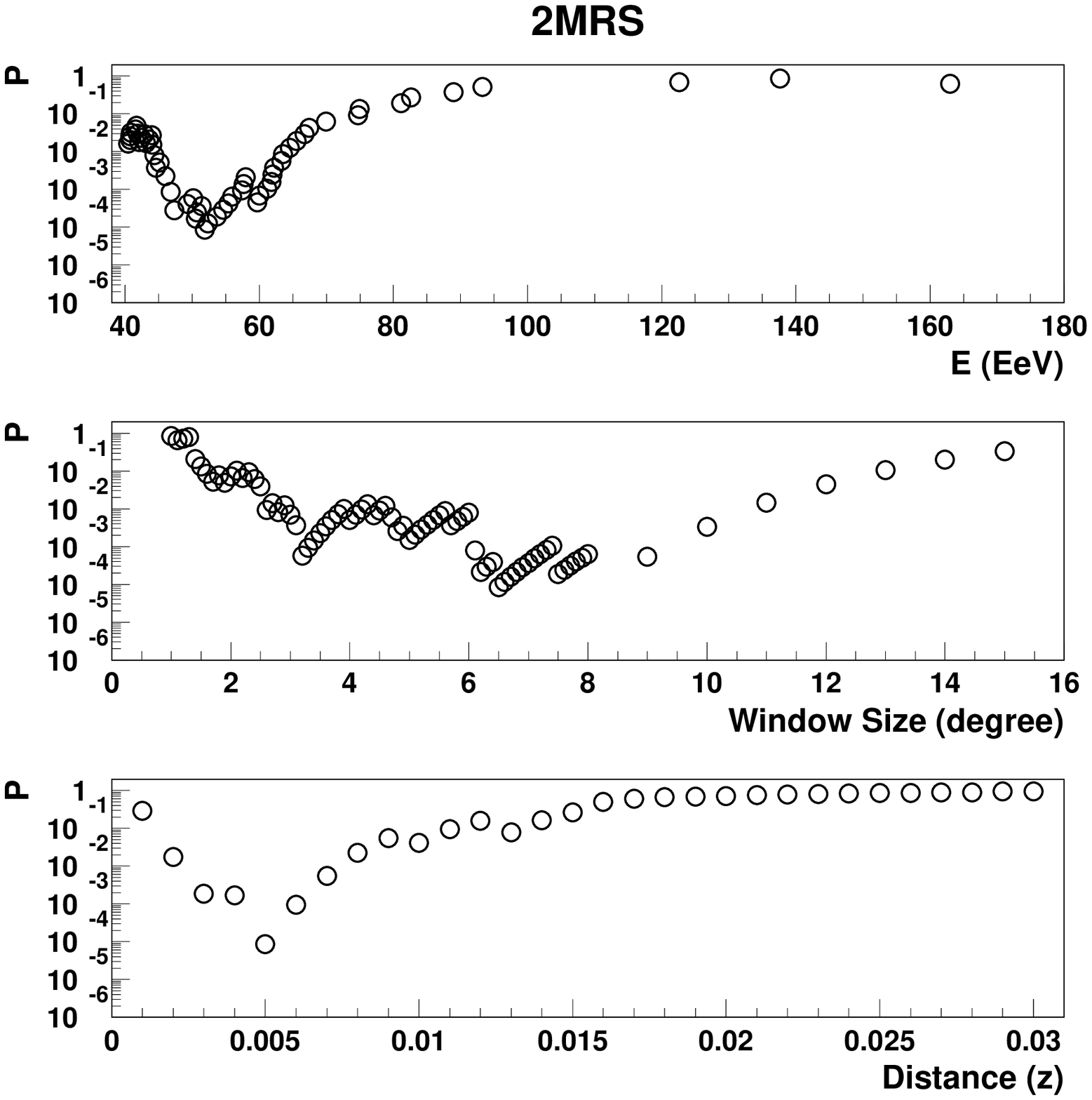}
\includegraphics[angle=0,keepaspectratio,scale=0.4]{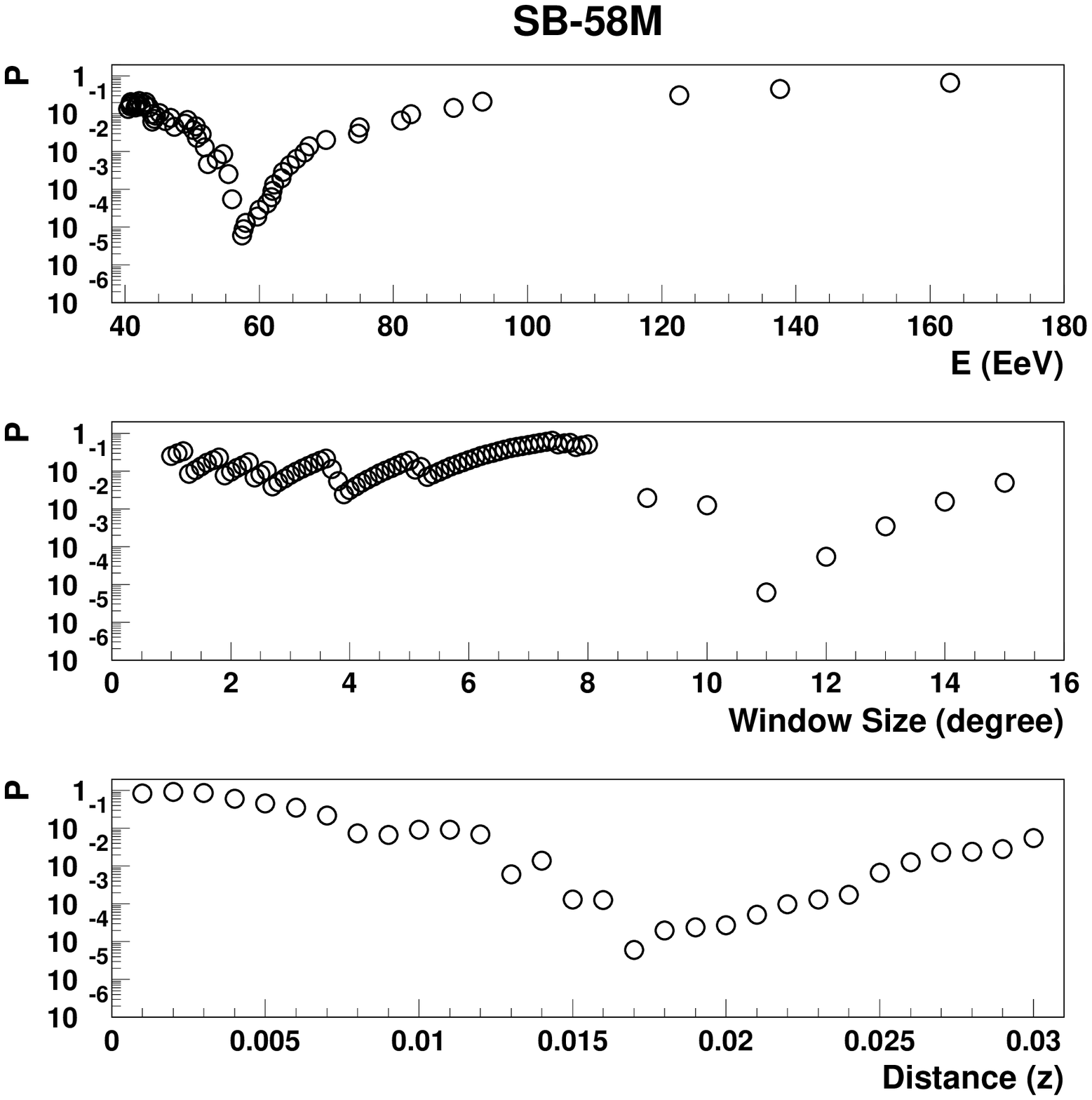}
\end{center}
\caption{Cumulative binomial probability distribution for the 3CRR 
(top left panel), VCV catalog (top right panel), 2MRS (bottom left panel), and SB-58M (bottom right panel).
Each panel shows the probability distribution with Energy threshold ($E_{\rm min}$) of 
observed cosmic rays (top), window $\psi$ (middle), redshift $z_{\rm max}$ (bottom).
In the each plot, the other two parameters are fixed at 
which parameter set provides $P_{\rm min}$.}
\label{a2}
\end{figure}
\begin{figure}
\begin{center}
\includegraphics[angle=0,keepaspectratio,scale=0.4]{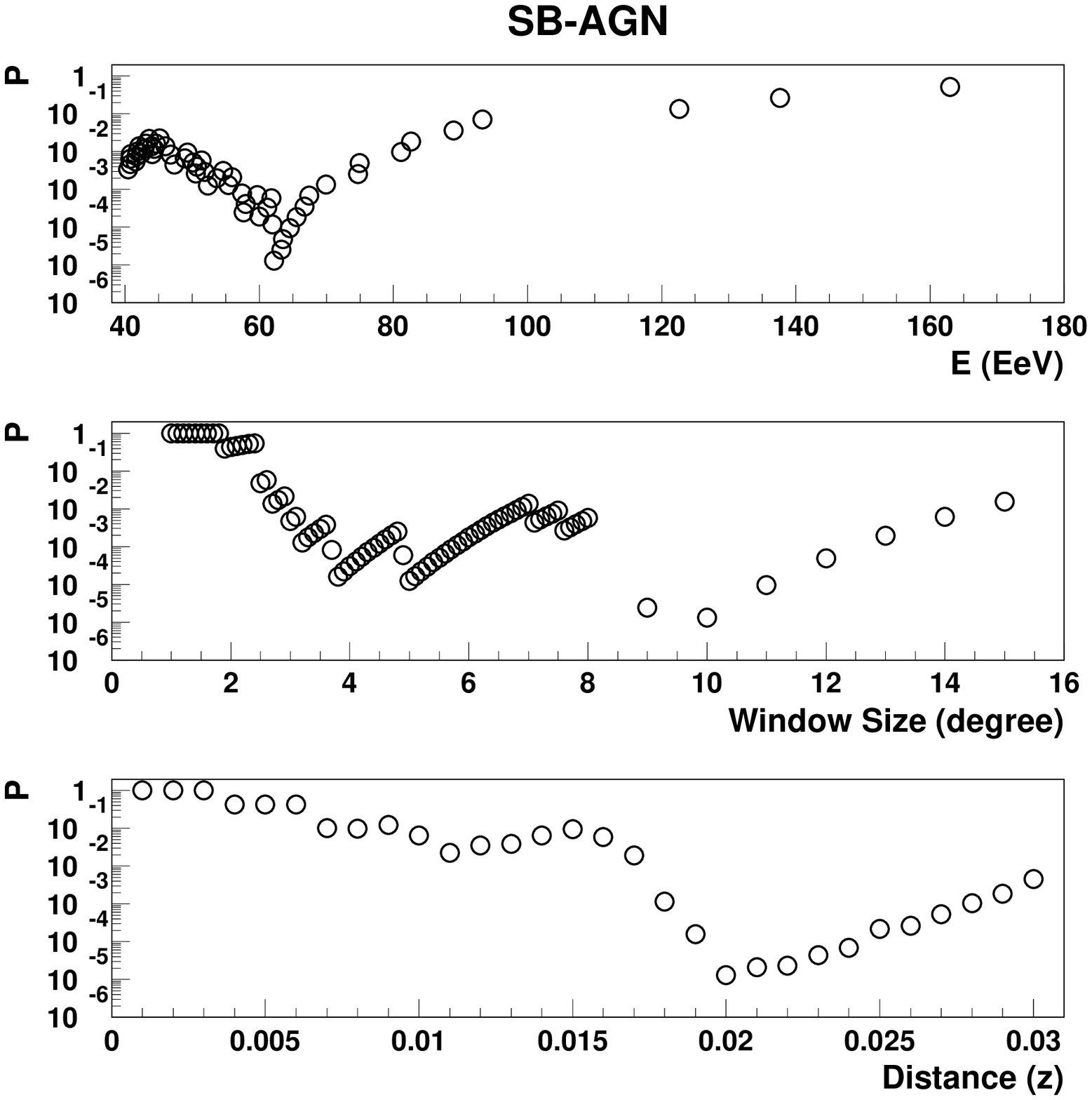}
\includegraphics[angle=0,keepaspectratio,scale=0.4]{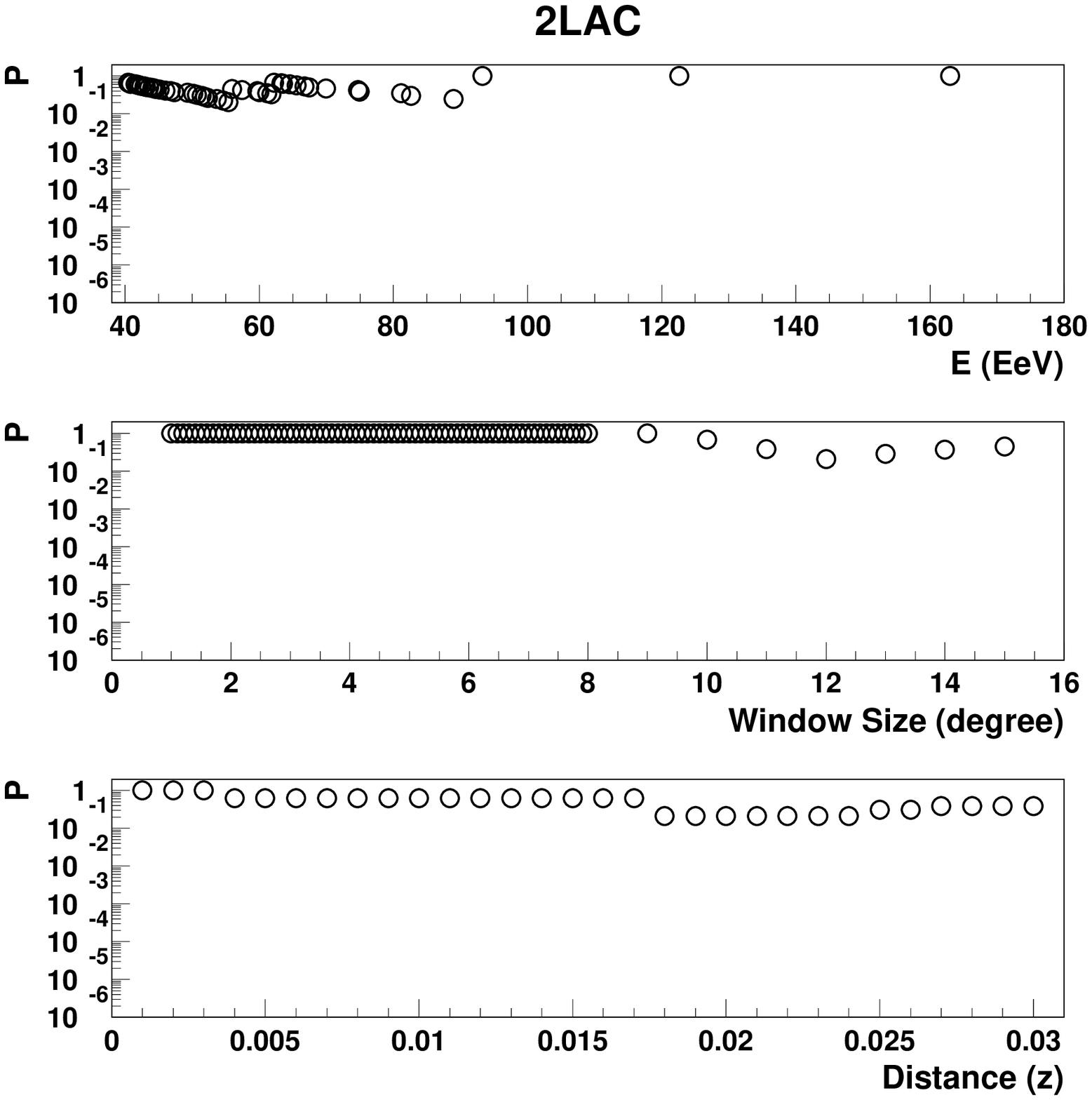}
\end{center}
\caption{Cumulative binomial probability distribution for the SB-AGN
(left panel), 2LAC (right panel).
Each panel shows the probability distribution with Energy threshold ($E_{\rm min}$) of 
observed cosmic rays (top), window $\psi$ (middle), redshift $z_{\rm max}$ (bottom).
In the each plot, the other two parameters are fixed at the optimum value.}
\label{a3}
\end{figure}

The results of the parameter scan are listed in Table~\ref{ChanceProbability}.
The smallest value of $P_{\rm min}^{\rm obs}$ among all the catalogs is 1.3$\times10^{-5}$ found in
the SB-AGN catalog with the best parameters
$(E_{\rm min}, \psi, z_{\rm max})_{\rm best} = (62.20~{\rm EeV}, 10^{\circ}, 0.020)$.
A sky map of the TA events and the objects under the condition $(E_{\rm min}, \psi, z_{\rm max})_{\rm best}$ 
 which gives the smallest $P_{\rm min}^{\rm obs}$ is shown in Figure~\ref{swiftbatagn}. 
All the observed UHECRs with $E \ge E_{\rm min}$ correlate with at least 
one object with $z \le z_{\rm max}$ in the SB-AGN catalog.
Figure~\ref{a2}, and~\ref{a3} show the probability as a function of each parameter $(E_{\rm min}, \psi, z_{\rm max})_{\rm best}$
while fixing the values of the other two at the optimum value for this data comparison.
%

\begin{figure}
\begin{center}
\includegraphics[angle=0,keepaspectratio,scale=0.6]{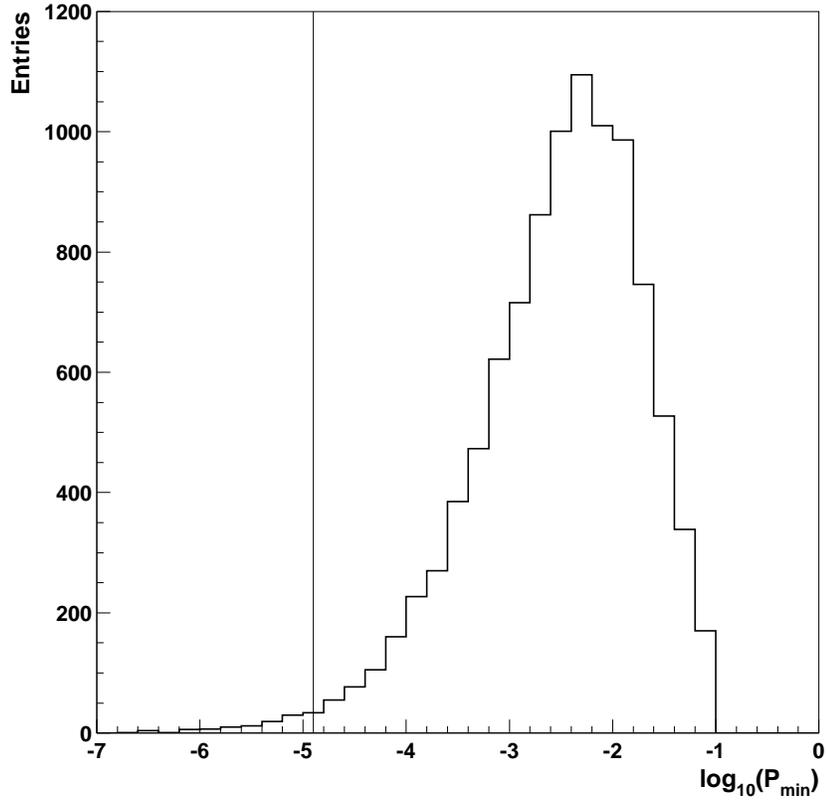}
\end{center}
\caption{Distribution of probability $P_{\rm min}^{\rm sim}$ for SB-AGN catalog 
determined from 10$^4$ simulated isotropic data sets. The
observed $P_{\rm min}^{\rm obs}=1.3\times 10^{-5}$ is shown as a vertical line.}
\label{b1}
\end{figure}

Now let us consider the penalty for the parameter scanning.
We evaluate the probability, $P_{\rm PPS}$, of finding a correlation
by chance with $P_{\rm min}^{\rm sim}$ smaller than that obtained from the 
data as follows~\citep[for a more detailed description of the penalty calculation  see, e.g.,][]{penalties}.
We generate 10$^4$ random sets of $N$ ``cosmic ray events'', where $N$ is the
same as the number of the observed events with energies greater than $40$~EeV.
For each of the mock event sets, the parameter scanning was carried out using
exactly same method as for the observed data set, and $P_{\rm min}^{\rm sim}$ was calculated.
Note that the parameters $(E_{\rm min}, \psi, z_{\rm max})_{\rm best}$ 
which yield $P_{\rm min}^{\rm sim}$ are different for each of the 10$^4$ trials.
The distribution of $P_{\rm min}^{\rm sim}$ in case of the SB-AGN catalog
is shown in Figure~\ref{b1} together with $P_{\rm min}^{\rm obs}$.
One can see that rather small values $P_{\rm min} \le 1.3 \times 10^{-5}$ 
can happen even though the simulated UHECR distribution is isotropic.

If we repeat the same experiment and the parameter scanning many times, 
the value of our result $P_{\rm min}^{\rm obs} = 1.3 \times 10^{-5}$ could be just
a chance occurrence. 
The probability including the Penalty for the Parameter Scanning (PPS)
is evaluated as $P_{\rm PPS} = 0.01$ for the SB-AGN
catalog, and the values for all the catalogs are listed in Table~\ref{ChanceProbability}. 
The smallest value of $P_{\rm PPS}$ among the catalogs is $0.01$ from 
the the SB-AGN catalog. This does not yet include the penalty for searching in several catalogs. 

If we have several catalogs, regardless of whether they are
independent or partially overlapping, there is a 
possibility of finding a catalog which gives the same 
or smaller $P_{\rm min}^{\rm obs}$ value by chance, 
even though there are no correlations between the events 
and the objects. 
The straightforward way to calculate the penalty factor associated 
with the partially overlapping catalogs, 
as is the case in our analysis, is to include all the 
catalogs in the Monte-Carlo simulation. 
Therefore, we have repeated the simulation with 10$^4$ mock sets 
as described above, but with the scanning performed in all six catalogs. 
Calculating the fraction of mock sets that show equal 
or better correlation than the data, we find that the final
%
probability with a Penalty of Parameter Scanning (PPS) and 
a Penalty of multi-Catalog Scanning (PCS) $P_{\rm PPS+PCS} = 0.09$.
Therefore, we conclude that no 
significant correlation between UHECRs and the 
astronomical objects is found in the current TA data set.

\subsection{UNCERTAINTIES}

First, we consider the effect of finite resolution in the scanning parameters.
The uncertainty in determination of the arrival directions and energy 
only make correlations worse due to direction smearing
and the contamination of the lower energy events than $E_{\rm min}$. 
Therefore, the obtained $P_{\rm min}^{\rm obs}$ already includes these resolution effects. 
The same concerns the uncertainty in the redshifts of the catalog objects.    

Consider now the effect of the systematic uncertainty in energy determination. 
As mentioned above, this uncertainty is 
$22\%$ \citep{tsunesada}.  
Note, however, that 
the present analysis with the parameter 
scanning is independent of absolute energy scale:  
the {\it energies} of the events are no more than 
keys for event sorting, 
and a systematic energy shift does not affect
the scanning in $E_{\rm min}$, hence the number
of events involved in the correlation with the objects
and the probability $P_{\rm min}^{\rm obs}$. 

The last issue to discuss is the incompleteness of the catalogs, 
which remains even after we cut out the regions around the Galactic plane.  
The objects in the VCV catalog are inhomogeneous because it is a mere compilation 
of objects detected under different conditions.
The completeness of the other catalogs, in particular the 2LAC,
could be affected by our cuts, particularly by the selection of objects with the known redshift. 
While the incompleteness may make the 
{\it interpretation}
of correlations ambiguous 
if they are present, it does not affect the calculation of $P_{\rm min}^{\rm obs}$. In fact, 
the effect of the incompleteness 
cancels out in $P_{\rm min}^{\rm obs}$ since the same set of objects is used to 
cross-correlate with the data and each mock event set. 
Therefore, the incompleteness of the catalogs
cannot produce spurious correlations (although it may, in principle, be responsible for their absence). 


\section{DISCUSSION}\label{discussion}

\subsection{SEARCH FOR CORRELATIONS WITH A SPECIFIC TYPE OF OBJECT}

So far we have treated the objects in each catalog
equally regardless of their class. 
Now let us examine whether there is a specific type 
of object that has stronger correlations with UHECRs 
than others. 
We will consider the case of the SB-AGN catalog which shows strongest correlations with UHECRs.

First, we count the number of objects of each class in the 
TA FOV with redshifts smaller than 0.02. 
%
Some of the objects are labeled ``unclassified" 
in the catalog. For these we used the information from
other surveys~\citep[][]{Unclassified1,Unclassified2,VCV,SwiftBAT}. 
The fractions of Seyfert 2, 1, 1.5, 1.9 and LINER galaxies 
in the SB-AGN catalog satisfying the above conditions are
0.441, 0.235, 0.132, 0.044, and 0.044, respectively (the total fraction of 
other class AGNs: 0.044, and the fraction of the unclassified AGN: 0.059). 
%
The total number of AGNs which are correlated with 
UHECRs is 22 with the parameters in Table~\ref{ChanceProbability}
(note that the number of UHECR events and that of 
AGNs are not the same because some of the events fall within the 
given angular distance from several sources). 
Among these 22 AGNs the fractions of Seyfert 
2, 1, 1.5, 1.9, LINER, and unclassified galaxies are 
0.455, 0.182, 0.227, 0.045, 0.045, and 0.045, respectively. 
We see that the largest difference is for the Seyfert 1.5 galaxies. 
The probability, $P$, of finding 5 or more correlated Seyfert 1.5 galaxies 
out of 22 by chance can be evaluated by the cumulative 
binomial probability with an expectation of 0.13, and is $P = 0.16$. 
%
Therefore,  no significant correlation with a 
specific type of AGN in the SB-AGN catalog is found.

%
%
%
%
%
%
%

\section{CONCLUSION}~\label{Conclusion}

We examine the correlations between the observed UHECR arrival directions
and the extragalactic objects from the different survey catalogs 
under assumption that the sources have the same intrinsic UHECR luminosities.
We use the TA-SD events with energies greater than 40~EeV
obtained in the first 40 months of  observation.
We search for maximum correlations by scanning over three parameters 
$E_{\rm min}$, $\psi$, and $z_{\rm max}$ in six different catalogs. 
%
The smallest chance probability among these six catalogs was found with
 the Swift BAT (60-month) AGN catalog,
$P_{\rm min}^{\rm obs}=1.3\times10^{-5}$.
This probability increases to $P_{\rm PPS}=0.01$ when we include the penalty 
for the three-parameter scanning in the Swift BAT catalog alone, and to 
$P_{\rm PPS+PCS} = 0.09$ when scanning in all the catalogs is taken into account.
%
%
Therefore, we conclude that no significant correlation
with the considered catalogs of extragalactic objects is found in the present TA data set.
%
%
%
Investigating specifically the case of the Swift BAT (60-month) AGN 
catalog which gives the strongest correlation, we find that 
no particular subclass of objects is responsible for this correlation.

\acknowledgments
The Telescope Array experiment is supported 
by the Japan Society for the Promotion of Science through
Grants-in-Aid for Scientific Research on Specially Promoted Research (21000002) 
``Extreme Phenomena in the Universe Explored by Highest Energy Cosmic Rays'', 
and the Inter-University Research Program of the Institute for Cosmic Ray 
Research;
by the U.S. National Science Foundation awards PHY-0307098, 
PHY-0601915, PHY-0703893, PHY-0758342, PHY-0848320, PHY-1069280, 
and PHY-1069286 (Utah) and 
PHY-0649681 (Rutgers); 
by the National Research Foundation of Korea 
(2006-0050031, 2007-0056005, 2007-0093860, 2010-0011378, 2010-0028071, R32-10130);
by the Russian Academy of Sciences, RFBR
grants 10-02-01406a and 11-02-01528a (INR),
IISN project No. 4.4509.10 and 
Belgian Science Policy under IUAP VI/11 (ULB).
The foundations of Dr. Ezekiel R. and Edna Wattis Dumke,
Willard L. Eccles and the George S. and Dolores Dore Eccles
all helped with generous donations. 
The State of Utah supported the project through its Economic Development
Board, and the University of Utah through the 
Office of the Vice President for Research. 
The experimental site became available through the cooperation of the 
Utah School and Institutional Trust Lands Administration (SITLA), 
U.S.~Bureau of Land Management and the U.S.~Air Force. 
We also wish to thank the people and the officials of Millard County,
Utah, for their steadfast and warm support. 
We gratefully acknowledge the contributions from the 
technical staffs of our home institutions as well as 
the University of Utah Center for High Performance Computing (CHPC).

\clearpage

\begin{table}
\begin{center}
\begin{tabular}{ccc}\hline
Parameter & Range & Step size  \\ \hline
Energy (EeV) & $E$$\ge$40 & Energy of each event by sorted order \\ 
Redshift (z) & 0.001$\le$ $z$ $\le$0.030 & 0.001 \\ 
\multirow{2}{*}{Window (degree)} & 1$\le$ $\psi$ $<$8 & 0.1 \\
                       & 8$\le$ $\psi$ $\le$15 & 1 \\ \hline
\end{tabular}
\caption{List of the scan regions and step size for each scan parameter.}\label{parameter_table}
\end{center}
\end{table}

\clearpage


\begin{table}
\begin{center}
\begin{tabular}{llcc}\hline
Catalog & Range & $N_{\rm all}$ & $N_{\rm target}$ \\ \hline
3CRR& compilation of Radio surveys & 173 & 16\\ 
2MRS & IR (1$-$2$\mu$m)& 43533 & 13547\\ 
SB-58M  & X-ray ($14-195$~keV)&1092& 161\\ 
SB-AGN & X-ray ($15-55$~keV)&428& 102\\ 
2LAC & $\gamma$-ray (100~MeV$-$100~GeV)&1126& 6\\ 
VCV & compilation of AGNs &168941& 762\\\hline
\end{tabular}
\caption{List of the configuration of the used catalogs. 
N$_{\rm all}$: number of all objects contained within the catalog,
N$_{\rm target}$: number of objects with the redshift $z<0.03$ within the TA FOV.}\label{catalog_table}
\end{center}
\end{table}


\clearpage

\begin{table}
\begin{center}
\begin{tabular}{llc}\hline
Catalog & Cut region (degree)& $N$ ($E\ge$40~EeV)  \\ \hline
3CRR & $|b|<$10$^{\circ}$, $\delta<$10$^{\circ}$  & 41\\ 
\multirow{2}{*}{2MRS}  & $|b|<$5$^{\circ}$ for 30$^{\circ}$ $\le l \le$ 330$^{\circ}$& \multirow{2}{*}{56}\\ 
 & $|b|<$8$^{\circ}$ otherwise & \\ 
SB-58M & None & 57 \\
SB-AGN & None & 57 \\
2LAC & $|b|<$10$^{\circ}$ & 49 \\ 
VCV & None & 57 \\ \hline
\end{tabular}
\caption{List of the cut region away from the Galactic plane of the each catalog and 
the number ($N$) of events remaining (the maximum number is 57). 
Symbols mean: $b$: Galactic latitude, $l$: Galactic longitude, $\delta$: declination of 
the equatorial coordinate.}\label{arrivaldirection_cut}
\end{center}
\end{table}

\clearpage


\begin{table}
\begin{center}
\begin{tabular}{ccccccccccc}\hline
Catalog & $E_{\rm min}$ & $\psi$ & $z_{\rm max}$ & $A$& $N$ & $k$ & $p$ &$P_{\rm min}$ &$P_{\rm PPS}$ \\ 
 & (EeV)& (degree)&$(z)$&  & & & & \\ \hline
3CRR& 66.77 & 2.0 & 0.017 &4&11 & 1 & 0.0020 & 2.2$\times$10$^{-2}$ & 0.75\\ 
2MRS & 51.85 & 6.5 & 0.005 &660&31 & 29 &0.62& 8.5$\times$10$^{-5}$ & 0.21\\ 
SB-58M & 57.46 & 11 & 0.017 &79& 25 & 25 & 0.68 & 6.1$\times$10$^{-5}$ & 0.04\\ 
SB-AGN &  62.20 & 10 & 0.020 &58& 17& 17 &0.52& 1.3$\times$10$^{-5}$ & 0.01\\ 
2LAC& 55.41 & 12 & 0.018 &3& 23& 3 &0.069& 2.1$\times$10$^{-1}$ & 0.83\\ 
VCV& 62.20 & 2.1 & 0.016 &288& 17& 8 &0.14& 8.6$\times$10$^{-4}$ & 0.25\\ \hline
\end{tabular}
\caption{Summary of correlations with the best parameter set (minimum threshold, Window size, maximum redshift) for each catalog.
$A$: number of objects with the redshift $\le z_{\rm max}$, 
$N$: number of observed cosmic ray events with the energy $E \ge E_{\rm min}$,
$k$: number of events correlated with objects,
$p$: probability of correlation for a single event from an isotropic distribution,
$P_{\rm min}$: the cumulative binomial probability to obtain $k$ or more estimated 
from an isotropic distribution, 
$P_{\rm PPS}$: the probability after including the penalties from parameter scanning.
}\label{ChanceProbability}
\end{center}
\end{table}


\clearpage

\clearpage

\clearpage

\clearpage

\clearpage

%

%

\end{document}